\documentclass[prl,twoside,amsfonts,amssymb,amsmath,titlepage,floatfix,letterpaper,twocolumn,superscriptaddress]{revtex4}
\usepackage{amsmath}
\usepackage{amsfonts}
\usepackage{amssymb}
\usepackage{graphicx}
\usepackage{graphicx,array}
\usepackage{dcolumn}

\usepackage{bm}

\newcommand{\bra}[1]{\ensuremath{\langle #1 \vert}}
\newcommand{\ket}[1]{\ensuremath{\vert #1 \rangle}}

\newcommand{\comment}[1]{}

\begin{document}

\title{Strong coupling between single photons in semiconductor microcavities}

\author{William T.M. Irvine}
\email[corresponding author: ]{william@physics.ucsb.edu}
\affiliation{Department of Physics, University of
California, Santa Barbara, CA 93106, USA}

\author{Kevin Hennessy}
\affiliation{ECE Department, University of
California, Santa Barbara, CA 93106, USA}

\author{Dirk Bouwmeester}
\affiliation{Department of Physics, University of
California, Santa Barbara, CA 93106, USA}

\date{\today}

\begin{abstract}
We discuss the observability of strong coupling between single photons in semiconductor microcavities  coupled by a $\chi^{(2)}$ nonlinearity. We present two schemes and analyze the feasibility of their practical implementation in three systems: photonic crystal defects, micropillars and microdisks,  fabricated out of  GaAs. We show that if  a weak coherent state is used to enhance the $\chi^{(2)}$ interaction, the strong coupling regime between two modes at different frequencies occupied by a {\it single} photon is within reach of current technology. The unstimulated  strong coupling of a single photon and a photon pair is very challenging and will require an improvement in mirocavity quality factors  of 2-4 orders of magnitude to be observable. 
\end{abstract}

\maketitle

The experimental realisations of strong coupling between  a single mode of an optical  cavity and a single atom have made it possible to demonstrate striking predictions of cavity quantum electrodynamics (QED)~\cite{strpaps}. Quantum information science has since provided motivation for gaining additional control of such strongly and coherently coupled systems.
Quantum dots embedded in  monolithic optical cavities have emerged as a promising system for the scalable implementation of cavity QED. Motivated in part  by this promise, a large effort has been put into the fabrication of small high quality monolithic microcavities~\cite{microc}.

In parallel, a  research effort  has begun  to make use of the high nonlinearities of  semiconductor materials such as GaAs to perform classical frequency conversion, using  microcavities to enhance the  electric field strength  and microstructures to provide the necessary phase-matching conditions~\cite{doubleres}.  Recently this approach was extended to  parametric down-conversion in nonlinear photonic crystals for the generation of  entangled photon pairs~\cite{us}.

Here we discuss the observability of strong coupling {\it between  single photons}  in  microcavities coupled by an optical nonlinearity, with an emphasis on the implementation  in realistic structures.

We consider two schemes: the first consists of two spatially overlapping single-mode cavities or a doubly-resonant cavity at frequencies $\omega_a$ and $\omega_b$ such that $\omega_a=2\omega_b$,  coupled by a $\chi^{(2)}$ non-linearity that  mediates the conversion of a photon in cavity $a$ to two photons in cavity $b$ and vice-versa. The second consists of three overlapping microcavities with frequencies $\omega_{a,b,c}$ satisfying $\omega_a=\omega_b+\omega_c$, with cavity c taken to be occupied by a coherent state $\ket{\alpha}_c$.
The effective nonlinearity in this case couples the conversion of a single photon in cavity $a$ to a single photon in cavity $b$ and is enhanced by the coherent state in mode $c$.

The dynamics of the two systems are similar. For the sake of clarity we will therefore solve the dynamics of the two-mode system and then state the corresponding results for the three mode system. 
 The Hamiltonian for the two-mode system is given by:
\begin{equation}
\hat{\mathcal{H}}=\hbar \omega_a \hat{a}^\dag \hat{a} + \hbar \omega_b \hat{b}^\dag \hat{b} + \hbar \Omega \big(\hat{a}(\hat{b}^{\dag})^2+ \hat{a}^\dag\hat{b}^2\big).
\label{eq:ham}
\end{equation}
where $\hat{a}^\dag(\hat{a}), \hat{b}^\dag(\hat{b})$ represent creation(annihilation) operators for modes $a,b$ and $\Omega$ is the strength of the coupling between the modes:
\begin{eqnarray}
\hbar \Omega & = &   \epsilon_0 \sqrt{\frac{\hbar \omega_a}{2\epsilon_0n_a^2V_a}} \frac{\hbar \omega_b}{2\epsilon_0n_b^2V_b}\nonumber \\
&&\times \int\mathrm{d}V \chi^{(2)}_{ijk}(r) E^i_a(r) E^j_b(r) E^k_b(r)
\label{eq:chi2eff}
\end{eqnarray}
where $\chi^{(2)}_{ijk}(r)$ is the non-linear susceptibility tensor, $\bm{E}_{a,b}(r)$ represent the spatial part of cavity modes $a,b$, normalised so that their maximum value is 1, $V_{a,b}$ represent the mode volumes defined as in Ref.~\cite{vmode} and we have adopted the repeated index summation convention.

Restricting our attention to the subspace:
$\ket{a}=\ket{1}_{a}\ket{0}_b$,  $\ket{b}=\ket{0}_a\ket{2}_b$.
we obtain the following  Hamiltonian:
\begin{equation}
\hat{H}=\hbar \left(\begin{array}{cc}
 \omega_a & \sqrt{2} \Omega e^{i\Delta t}\\
\sqrt{2}\Omega e^{-i \Delta t} & 2  \omega_b
\end{array}\right)
\end{equation}
where $\Delta=\omega_a-2\omega_b$ is the detuning between the cavities.
This is the well known Hamiltonian for  two  states \ket{a} and \ket{b} coupled by an interaction $\hbar \sqrt{2} \Omega(\ket{a}\bra{b} e^{i\Delta t} + \ket{b}\bra{a} e^{-i\Delta t})$ first discussed by Rabi~\cite{rabi}. It is   analogous to the Jaynes-Cummings Hamiltonian for  an atomic transition coupled to a single cavity mode~\cite{jc}, with the role of the excited atom played by the two photons in mode $b$, the role of the cavity photon played by the photon in mode $a$ and the role of the minimal  coupling played by $\hbar \sqrt{2} \Omega$.  The eigen-states are time-dependent superpositions of  the uncoupled eigen-states \ket{a} and \ket{b}, with  energies given by:
\begin{equation}
E_{\pm}= \frac{\hbar}{2} (\omega_a+2\omega_b)\pm \frac{\hbar}{2} \sqrt{(2\sqrt{2} \Omega)^2+\Delta^2}
\label{eq:split}
\end{equation}
 Just as in the atom-cavity case, if the system is prepared in one state, say \ket{a}, and $\Delta=0$, the time evolution will consist of Rabi-flopping  between states \ket{a} and \ket{b} at twice the Rabi frequency   $2\Omega_\mathrm{R} = 2\sqrt{2} \Omega$.

In practice,  the bare cavities  will leak photons at a rate that will depend  on the details of the particular cavity. The possibility of observing an oscillation  will depend on the ratio of the period  of a Rabi oscillation to the cavity decay time.
In the context of atom-cavity systems, if the oscillation is in principle observable,
 the system   is said to be in the strong coupling regime.  A precise criterion for the discussion of strong coupling in the system presented here, is afforded by solutions of the following master equation~\cite{agarwal}:
\begin{eqnarray}
\dot{\hat{\rho}}=-\frac{i}{\hbar}\big[\hat{H}_{\mathrm{int}},\hat{\rho}\big]&-&\frac{1}{2\tau_a}\big(\hat{a}^\dag\hat{a}\hat{\rho}+\hat{\rho}\hat{a}^\dag\hat{a}\big)+\frac{1}{\tau_a}\hat{a}\hat{\rho}\hat{a}^\dag \nonumber \\
&-&\frac{1}{2\tau_b}\big(\hat{b}^\dag\hat{b}\hat{\rho}+\hat{\rho}\hat{b}^\dag\hat{b}\big)+\frac{1}{\tau_b}\hat{b}\hat{\rho}\hat{b}^\dag,
\label{eq:master}
\end{eqnarray}
where $\hat{\rho}$ represents the reduced density matrix for the two cavities,  $\hat{H}_{\mathrm{int}}$ represents the interaction part of the Hamiltonian of Eq.~\ref{eq:ham} and the second and third terms model the loss of photons from cavities $a$ and $b$.
If the system is prepared in the state $\ket{1}_a\ket{0}_b$, the four  joint states of the cavities relevant to the time evolution are:
$\ket{1}=\ket{1}_a\ket{0}_b$,
$\ket{2}=\ket{0}_a\ket{2}_b$,
$\ket{3}=\ket{0}_a\ket{1}_b$ and
$\ket{4}=\ket{0}_a\ket{0}_b$.
Expressing $\hat{\rho}$ in this basis, and writing out the master equation for each component separately, the following closed subset can be found:
\begin{equation}
\left(\begin{array}{c}
\dot{\rho}_{11} \\
\dot{\rho}_{22} \\
\dot{V}
\end{array}\right)=\left(\begin{array}{ccc} -\frac{1}{\tau_a} & 0 & i\sqrt{2}\Omega  \\  0 & -\frac{2}{\tau_b}  & -i\sqrt{2}\Omega \\ i2\sqrt{2}\Omega  & -i2\sqrt{2}\Omega & -(\frac{1}{\tau_b}+\frac{1}{2\tau_a}) \end{array}\right)\left(\begin{array}{c}
\rho_{11} \\
\rho_{22} \\
V
\end{array}\right)
\end{equation}
where $\rho_{ij}=\bra{i}\rho\ket{j}$ and $V=\rho_{12}-\rho_{21}$. The matrix elements   $\rho_{33}$ and $\rho_{44}$ are determined in turn by: $\dot{\rho}_{33} =\frac{2}{\tau_b}\rho_{22}  - \frac{1}{\tau_b}\rho_{33}$
 and $\dot{\rho}_{44} =\frac{1}{\tau_a}\rho_{11}  +\frac{1}{\tau_b}\rho_{33} $.

\begin{figure}
\includegraphics[width= \columnwidth]{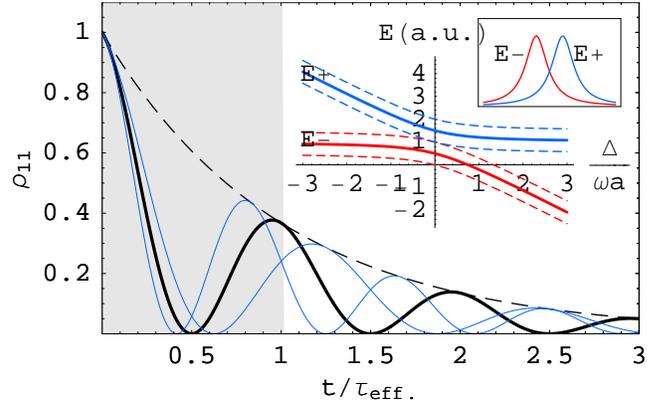}
\caption{Evolution of $\rho_{11}$ for initial conditions $\rho_{11}=1,\rho_{22}=V=0$. The frequency of the oscillation is $2\Omega_{\mathrm{R}}=2\sqrt{2}\Omega$ corresponding to a period $\tau_{\mathrm{R}}=2\pi/\Omega_{\mathrm{R}}$. The $1/e$ time of the decay is given by $\tau_{\mathrm{eff.}}=(1/\tau_a+1/2\tau_b)^{-1}$.  The strong coupling criterion proposed here consists of requiring that the revival of the oscillation takes place before the $1/e$ time of the decay. Inset: Solid lines: $\mathrm{E}_\pm(\Delta)$ for fixed $\omega_a$. Dashed lines: FWHM of $\mathrm{E}_\pm$. The energy splitting at zero detuning $\Delta=0$ provides a second, weaker  criterion for strong coupling. }
\label{fig:time}
\end{figure}

The eigen-values of the matrix tell us whether the solutions 
have the character of a damped oscillation, or a critically damped
exponential decay. They are given by:
\begin{eqnarray}
\lambda_0 & = & -(\frac{1}{2\tau_a}+\frac{1}{\tau_b})\\
\lambda_{\pm}&=& -(\frac{1}{2\tau_a}+\frac{1}{\tau_b}) \pm \sqrt{\Big(\frac{1}{2\tau_a}-\frac{1}{\tau_b}\Big)^2-(2\sqrt{2}\Omega)^2} \label{eq:lamdapm}
\end{eqnarray}
The time evolution of two of the eigen-states of the matrix will be oscillatory if $2\sqrt{2}\Omega>\vert\frac{\tau_b-2\tau_a}{2\tau_a\tau_b}\vert$. The frequency of the oscillation will be $2\Omega_{\mathrm{R}}=\sqrt{(2\sqrt{2}\Omega)^2-(\frac{1}{2\tau_a}-\frac{1}{\tau_b})^2}$ and the $1/e$ time of the decay of the oscillation is
$\frac{1}{\tau_{\mathrm{eff.}}}=\frac{1}{2\tau_a}+\frac{1}{\tau_b}$.
In the context of the atom-cavity system, oscillatory behaviour is synonymous with strong coupling.  In the present context, adoption of such a criterion would not
be as restrictive as required for it to be meaningful, since $\lambda_\pm$ has an imaginary part for $\tau_b=2\tau_a$ regardless of whether the Rabi period is at all comparable to the cavity decay time. Note that the definition in the atomic case is meaningful because the atomic lifetime is always much longer than the cavity lifetime, reducing the strong coupling condition to: $\sqrt{2}\Omega>\frac{1}{2}\vert\frac{1}{\tau_a}\vert$. We therefore suggest the following criterion for strong coupling in this system:
$\tau_{\mathrm{eff.}} \geq   \frac{\tau_{\mathrm{R}}}{2} = \frac{2\pi}{2\sqrt{2}\Omega} $
illustrated in Figure~\ref{fig:time}.

An alternative and somewhat less restrictive criterion is provided by the resolvability of the energy splitting (Eq.~\ref{eq:split}). 
The width of the split energy levels (Eq.~\ref{eq:split}) at
non-zero detuning can be obtained by evaluating the master
equation (Eq.~\ref{eq:master}) in the dressed state
basis and taking the Fourier transform of the resulting $e^{\lambda_\pm t}$~\cite{ftenergy,agarwal}. The result is:
$\Gamma_\pm^\mathrm{eff.}=\frac{\Gamma_a}{2}
\frac{\Omega'\pm\Delta}{2\Omega'}+\Gamma_b
\frac{\Omega'\mp\Delta}{2\Omega'}$ where
$\Omega'=\sqrt{(2\sqrt{2}\Omega)^2+\Delta^2}$.
Figure~\ref{fig:time} (inset) shows a plot of the energy
eigen-states as a function of detuning accompanied by the FWHM of
the lines obtained by this approach. Applying the Rayleigh
criterion (separation=FWHM) for the resolvability of the splitting
at zero detuning we obtain the following `spectral' strong
coupling criterion: $ \pi \tau_{\mathrm{eff.}} \geq
\frac{\tau_{\mathrm{R}}}{2} = \frac{2\pi}{2\sqrt{2}\Omega},$
illustrated in Figure~\ref{fig:time} (inset).

We now state the results of a similar calculation for the three mode system, which is governed by the following effective Hamiltonian:
\begin{equation}
\hat{\mathcal{H}}=\hbar \omega_a \hat{a}^\dag \hat{a} + \hbar \omega_b \hat{b}^\dag \hat{b} +  \hbar \omega_c \vert\alpha\vert^2 + \hbar \Omega \vert\alpha\vert  \big(\hat{a}\hat{b}^{\dag}+  \hat{a}^\dag\hat{b}\big),
\label{eq:hamm2}
\end{equation}
where $\hbar \Omega= \epsilon_0 \sqrt{\frac{\hbar \omega_a}{2\epsilon_0n_a^2V_a}} \sqrt{\frac{\hbar \omega_b}{2\epsilon_0n_b^2V_b}} \sqrt{\frac{\hbar \omega_c}{2\epsilon_0n_c^2V_c}} \nonumber \times \int\mathrm{d}V \chi^{(2)}_{ijk}(r) E^i_a(r) E^j_b(r) E^k_c(r)$, and $\vert\alpha\vert^2$ is  the mean photon number in cavity $c$.
The dynamics in  a similarly chosen subspace: 
$\ket{1}=\ket{1}_a\ket{0}_b$,
$\ket{2}=\ket{0}_a\ket{1}_b$,
$\ket{3}=\ket{0}_a\ket{0}_b$
are identical, with the effective decay rate replaced by $\tau_{\mathrm{eff.}}=(1/\tau_a+1/\tau_b)^{-1}$ and $\sqrt{2}\Omega$ of Eq.~\ref{eq:ham} replaced by $\vert \alpha \vert \Omega$ of Eq.~\ref{eq:hamm2}. It is important to note that mode $c$ need not be a high quality cavity mode; its main purpose is to enhance the nonlinear interaction between cavities $a$ and $b$ and provide the missing energy for the conversion. An implementation  of mode $c$ could be  a weakly confined laser beam impinging on modes $a$ and  $b$.

We now address the role of phase-matching, mode overlap and the tensor nature of $\chi^{(2)}$. 
In conventional frequency-conversion schemes, the requirement that photons interacting through a $\chi^{(2)}$ nonlinearity phase match, can be understood by inspection of the overlap integral for the fields (Eq.~\ref{eq:chi2eff} for example). In the case that the eigen-modes correspond to travelling waves, the $\bm{E}$'s will have the form of complex exponentials in the direction of travel.  The overlap  is then proportional to a sinc function, leading to the phase-matching requirement that $\Delta k\sim1/L$ where $L$ is the length of the path along which the photons interact. If, however, as is typically the case in the systems considered here, the modes take the form of standing waves, the overlap integral  simply takes the form of a spatial overlap of real field amplitudes. Phase matching thus does not play a role in the systems considered here, but in turn the design of  cavities with good overlap becomes of central importance.
The polarization of the modes also has to be taken into account,
since the $\chi^{(2)}$ interaction is tensorial. This is done by contracting  the electric field vectors with the $\chi^{(2)}$ tensor, before performing the
overlap integral. The effect of this on the value of the overlap integral
 depends on the detailed geometry of the system and on the
symmetry group of the non-linear material. We will consider, as an example, structures fabricated out of
GaAs which has crystalline structure of the $\bar{4}3m$ type.  This  has the following implications for the values of the  components of  $\chi^{(2)}$~\cite{boyd}: $ \chi^{(2)}_{xyz}=\chi^{(2)}_{xzy}=\chi^{(2)}_{yxz}=\chi^{(2)}_{yzx}=\chi^{(2)}_{zxy}=\chi^{(2)}_{zyx}= \vert \chi^{(2)} \vert $.
This makes it comparatively simple to orient the GaAs lattice so that the contraction at any one point  does not   lead to a reduction in the effective value of the non-linearity, for example, in the case of all three polarizations being  aligned and pointing in the (111) direction, the value of the contraction of the three polarization unit vectors with the $\bar{4}3m$ $\chi^{(2)}$ tensor is $\sim1.15  \vert \chi^{(2)} \vert$. The value of  $\vert \chi^{(2)} \vert$ in GaAs is 200pm/V at wavelengths of around $1.5\mu$m~\cite{chi2GaAs}, two orders of magnitude  greater than that of  common nonlinear crystals such as BBO~\cite{chi2nlcryst}. This high value of the nonlinearity, typical of semiconductor materials, 
 together  with the enhancement in the electric field per photon afforded by the microcavity  is what makes the proposed schemes viable.

We now turn to a discussion of three systems that could provide a setting for the schemes discussed above, all of which have  been used recently to study  strong coupling effects between photons and quantum dots~\cite{dotstr}; they are:  photonic crystal defect microcavities~\cite{firstpcdmc}, microdiscs~\cite{firstmicrod} and micropillars~\cite{firstmicrop}.

{\it Photonic crystal defect microcavities (PCDMC):} PCDMC modes  are confined modes that arise when one or more unit cells are removed from a photonic crystal that has a band gap at the relevant frequencies.
They offer  an unprecedented ability to control cavity mode volume, polarization and frequency.
They have been demonstrated in photonic crystals consisting of a two-dimensional periodic  lattice of holes etched in a thin membrane  of GaAs, with confinement  in the direction perpendicular to the plane of the periodicity provided by total internal reflection.
Cavities with  quality factors (Q) of up to 18000 (corresponding to confinement times of $\tau(\lambda=1\mu\mathrm{m})=9.5$ps) and mode volumes of  $0.7 (\lambda/n)^3$ have been demonstrated  at a wavelength of $1\mu$m~\cite{laser}.  In-and-out coupling can be achieved by integrating wave-guides within the photonic crystals~\cite{waveg}, through an optical fiber~\cite{fibre} or by free-space optics~\cite{laser}.

The design of  PCDMCs with multiple resonances  is challenging but does not seem unrealistic. A good starting point is the calculation of the band-gap maps of various defectless lattices, taking into account the finite thickness of the membrane. This is particularly  important since it can lead to a strong modification  and in some cases even the closing  of higher order band gaps~\cite{ita,itapriv}. Having found a pattern that has appropriate band gaps, one has to seek high quality defect modes by removing one or more holes.  An intuitively interesting class of lattices to pursue are those with two periodicities built into them, such as triangular lattices with a multiple atom unit cell of which Archimedean lattices~\cite{arch} are an example. An exciting possibility is also presented by Penrose tiling based photonic quasi-crystals, in which a single-frequency cavity has been recently demonstrated~\cite{pendefectdem} and which have been shown theoretically to support modes at widely differing resonant frequencies~\cite{quasiband,tbpe}. 

To estimate how close the strong coupling regime is to being achievable with current technology, we first consider the three mode scheme,  estimating the Rabi period   and comparing it to the cavity lifetime as follows: we assume  that a doubly-resonant PCDMC can be designed  that will support overlapping modes at frequencies $\omega_a$ and $\omega_b$ with Q's  similar to those obtained in Ref.~\cite{laser}.  Taking modes $a$, $b$ and $c$ to overlap well, all three polarizations to be the same (TE) and  the growth direction to be  (111), we estimate the overlap integral in Eq.~\ref{eq:chi2eff}  to be equal to $\frac{1}{2}  \vert \chi^{(2)}_{\mathrm{GaAs}} \vert V_a$. Taking $V_c=f_c V_a$, where a realistic range for $f_c$ is 1-100, an average of $n$ photons in mode $c$ and  $\lambda_b=1.5\mu$m  we then obtain an oscillation period $\tau_{\mathrm{R}}/2\sim5 \sqrt{\frac{f_c}{n}}$ns. The corresponding cavity effective lifetime is $\tau_{\mathrm{eff.}}=4.8$ps. The strong coupling regime is thus within reach with  an average of  $10^6 f_c$ photons in mode $c$. A similarly constructed estimate in the unseeded case yields $\tau_{\mathrm{R}}/2\sim18$ns, 3 orders of magnitude away according to the spectral criterion.

{\it Micro-pillars:} Micro-pillars  are microscopic cylinders etched out of closely spaced Bragg mirrors, with confinement in the radial direction  provided by  index contrast. They present clear in and out coupling advantages. The design of a doubly-resonant Bragg mirror configuration, which gives good mode overlap has been studied extensively for the case of parallel mirrors~\cite{doubleres} and is readily achievable. A cavity with a Q of 27700 ($\tau(\lambda=930\mathrm{nm})=13.6$ps) and a mode volume of $100(\lambda/n)^3$ was  demonstrated in Ref.~\cite{bestmicropillars} at a wavelength of 930nm. The corresponding periods are $\tau_{\mathrm{R}}/2\sim 44 \sqrt{\frac{f_c}{n}}$ns,  and $\tau_{\mathrm{eff.}}=8.0$ps. The strong coupling regime is thus within reach with  an average of  $3 \times 10^7 f_c$  photons in mode $c$; in the unseeded case ($\tau_{\mathrm{R}}/2\sim177$ns) it is 4 orders of magnitude away. 

{\it Micro-disk resonators:} Micro-disk resonators are resonators that consist of a thin disk of material, supported by a column. The high-Q modes correspond to whispering gallery modes that hug the outside walls of the resonator. Typically, defects in the microdisks couple counter propagating  modes to create standing wave modes~\cite{microdbest}. The in-and-out coupling can be achieved by use of a fiber~\cite{microdbest}. Q's of 360000 ($\tau(\lambda=1.4\mu\mathrm{m})=267.3$ps) have been demonstrated in GaAs at a wavelengths of 1.4$\mu$m~\cite{microdbest},   with a mode volume of $6(\lambda/n)^3$.  Making similar assumptions to those made for the PCDMCs we obtain $\tau_{\mathrm{R}}/2\sim 37 \sqrt{\frac{f_c}{n}}$ns,  and $\tau_{\mathrm{eff.}}=95$ps. The strong coupling regime is thus within reach with  an average of only $76 \times 10^3 f_c$ photons in mode $c$ whereas  in the unseeded case  ($\tau_{\mathrm{R}}/2\sim148$ns), it is only 2 orders of magnitude away.

As a final point we discuss possible schemes to measure the strong coupling effects presented here. In the spectral domain, one could measure the transmission of the cavities as a function of the detuning between the cavities or as a function of the coherent state intensity.  The latter is much simpler, but can only be implemented in the three mode scheme. 

In the time domain, one could initiate the coupled system with a photon in one of the modes, for example by sending an appropriately shaped pulse into one of the cavities, one could then wait and measure the photon emission from cavities $a$ and $b$ as a function of time. A simpler alternative that works in the three mode case, is to send a photon into cavity $a$ and then apply a  Rabi $\pi$-pulse through cavity $c$ to deterministically convert the photon in mode $a$ to a photon in mode $b$.

In conclusion we have  discussed the observability of strong coupling between  single photons  in semiconductor   microcavities coupled by an optical nonlinearity. We have shown that if the process is stimulated by a weak coherent state, the strong coupling regime is within reach of current technology. Engineering structures in which the unstimulated process could be observed appears to be a challenging goal for years to come. 
 The observation of such a coupling would constitute a new regime for photons in quantum optical systems. Aside from the design of structures optimised for the implementation of the  schemes presented here, an extension of the present work would be the investigation of ways to  further enhance the nonlinearities by engineering the material properties, ways to integrate sources, such as quantum dots, with the present schemes to create deterministic sources of entangled photon pairs and ways to 
  implement  quantum logic gates between strongly coupled single photons in both the two and three-mode schemes.

We acknowledge C. Simon and M. Rakher for useful discussions and financial support from NSF
grants 0304678 and DARPA/ARO.


\begin{thebibliography}{99}



\bibitem{strpaps}
J. Raimond, J.M., Brune, M. and Haroche, S. Rev.Mod. Phys. 73, 565 (2001);
J. McKeever, A. Boca, A.D. Boozer, J. R. Buck and  H. J. Kimble, Nature {\bf 425}, 268 (2003);
H. Mabuchi and A. C. Doherty,  Science {\bf 298}, 1372 (2002).

\bibitem{microc}
K.J.~Vahala, Nature {\bf 424}, 839 (2003).


\bibitem{doubleres}
M.~Liscidini and L.C.~Andreani, Appl. Phys. Lett. {\bf 85}, 1883 (2004); 
V.~Berger, X.~Marcadet and J.~Nagle, Pure Appl. Opt. {\bf 7}, 319 (1998); 
V.~Berger, J. Opt. Soc. Am. B {\bf 14}, 1351 (1997). 



\bibitem{us} M.J.A. de Dood, W. T. M. Irvine, and D. Bouwmeester, Phys. Rev. Lett. {\bf 93}, 040504 (2004); 
W.T.M. Irvine, M.J.A. de Dood, D. Bouwmeester, 
   Phys. Rev. A {\bf 72}, 043815 (2005). 


\bibitem{vmode}
L.C. Andreani, G. Panzarini and J-M. G\'erard, Phys. Rev. B {\bf 60}, 13276 (1999). 


\bibitem{rabi} I.I.~Rabi, Phys. Rev. {\bf 51}, 652 (1937). 

\bibitem{jc} E.T.~Jaynes and F.W.~Cummings, Proc. IEEE {\bf 51}, 89 (1963). 


\bibitem{agarwal} 
G.S. Agarwal,  Phys. Rev. Lett. {\bf 73}, 522 (1994).


\bibitem{ftenergy} 
H.J.~Carmichael {\it et al.},
Phys. Rev. A {\bf 40}, 5516 (1989).



\bibitem{boyd}
R.W.~Boyd, {\it Nonlinear optics}, Academic Press (2003).



\bibitem{chi2GaAs}
S. Bergfeld and W. Daum, Phys. Rev. Lett. Vol.{\bf 90}, 036801 (2003).



\bibitem{chi2nlcryst}
R.C.~Eckardt, H.~Masuda, Y.X.~Fan and R.L.  Byer,
IEEE J. Quant. Electron.,  Vol.{\bf  26},  p922  (1990).


\bibitem{dotstr} 
J. P. Reithmaier, {\it et. al.}, Nature {\bf 432}, 197 (2004); 
T.~Yoshie {\it et. al}, Nature {\bf 432}, 200 (2004).

\bibitem{firstpcdmc}
O.~Painter {\it et al.},
Science {\bf 284}, 1819 (1999). 

\bibitem{firstmicrod}
E.~Peter {\it et al.}, Phys. Rev. Lett. {\bf 95} 067401 (2005);
K.~Srinivasan {\it et al.},
Appl. Phys. Lett. {\bf 86}, 151106 (2005);
S.L.~McCall, {\it et al.},
Appl. Phys. Lett. {\bf 60}, 289 (1992).

\bibitem{firstmicrop}
J. M. G\'erard, {\it et al.}, Appl. Phys. Lett. {\bf 69}, 449 (1996);
G.S.~Solomon, M.~Pelton and Y.~Yamamoto, Phys. Status Solidi {\bf 178}, 341 (2000).



\bibitem{laser} S.~Strau\ss {\it et al.}, Submitted for publication.

\bibitem{waveg}
S.~Noda, A.~Chutinan and M.~Imada, Nature {\bf 407}, 608 (2000).

\bibitem{fibre}
P.E.~Barclay, K.~Srinivasan, M.~Borselli and O.~Painter,
Opt. Lett. {\bf 29}, 697 (2004).



\bibitem{ita}
L.C. Andreani and M. Agio, IEEE J. Quantum Electron. {\bf 38}, 891 (2002).

\bibitem{itapriv}
D.~Gerace and L.C. Andreani,  private communication.

\bibitem{arch}
M.~Rattier {\it et al.},
Appl. Phys. Lett. {\bf 83}, 1283 (2003);
Gr\"unbaum and Shephard, Tilings and Patterns, Freeman, New York (1987).


\bibitem{pendefectdem}
K.~Nozaki and T.~Baba,
Appl. Phys. Lett. {\bf 84}, 4875 (2004); 
S-K.~Kim {\it et al.}, 
Appl. Phys. Lett. {\bf 86} 031101 (2005).

\bibitem{tbpe} 
To be presented elsewhere.

\bibitem{quasiband}
Y.S.~Chan, C.T.~Chan and Z.Y.~Liu, Phys. Rev. Lett. {\bf 80}, 956 (1998).


\bibitem{bestmicropillars}
A.~L\"offler, J.P.~Reithmaier, G.~S\c{e}k, C.~Hoffmann, S.~Reitzenstein, M.~Kamp and A.~Forchel, Appl. Phys. Lett. {\bf 86}, 111105 (2005).



\bibitem{microdbest}
K.~Srinivasan, M.~Borselli, T.J.~Johnson, P.E.~Barclay and O.~Painter,
Applied Physics Letters, {\bf 86}, 151106 (2005).






\bibitem{microtoroid}
V.S.~Ilchenko, A.A.~Savchenkov, A.B. Matsko and L.~Maleki, Phys. Rev. Lett. {\bf 92}, 043903 (2004).



\bibitem{smallv}
J.T.~Robinson, C.~Manolatou, L.~Chen and M.~Lipson,
Phys. Rev. Lett. {\bf 95}, 143901 (2005).



\end{thebibliography}
\end{document}